\documentclass[aps,prl,showpacs,footinbib,superscriptaddress,twocolumn,longbibliography,floatfix,10pt]{revtex4-2}

\usepackage{amsmath,amssymb,bm,mathtools}
\usepackage{microtype}
\usepackage{upgreek}
\usepackage{xcolor}
\usepackage{graphicx}
\usepackage{enumitem}
\usepackage[colorlinks=true,citecolor=darkgreen,linkcolor=darkgreen,urlcolor=darkgreen]{hyperref}
\definecolor{darkgreen}{RGB}{0, 72, 0}

\newcommand{\C}{\mathbb C}
\newcommand{\PP}{\mathcal P}
\newcommand{\QQ}{\mathcal Q}
\newcommand{\GG}{\mathbb G}
\newcommand{\calG}{\mathcal G}
\newcommand{\calA}{\mathcal A}
\newcommand{\dd}{\mathrm d}
\newcommand{\ord}{\operatorname{ord}}
\newcommand{\abs}[1]{\left|#1\right|}
\newcommand{\wtp}{\widetilde p}
\newtheorem{theorem}{Theorem}

\newcommand{\placeholderbox}[2]{%
\fbox{\begin{minipage}[c][#1][c]{0.92\linewidth}
\centering\small #2
\end{minipage}}}
\newcommand{\missingfigure}[1]{\placeholderbox{1.18in}{#1}}
\newcommand{\maybegraphics}[3][]{%
\IfFileExists{#2}{\includegraphics[#1]{#2}}{\missingfigure{#3}}}

\begin{document}

\title{Non-Hermitian Edge State Endocytosis}

\author{Si-Yu Yuan}
\affiliation{Department of Physics, National University of Singapore, Singapore 117542}

\author{Wen-Tan Xue}
\affiliation{Department of Physics, National University of Singapore, Singapore 117542}

\author{Ching Hua Lee}
\email{phylch@nus.edu.sg}
\affiliation{Department of Physics, National University of Singapore, Singapore 117542}

\begin{abstract}
An isolated edge state observed in a finite open chain is usually expected to survive the thermodynamic limit (TDL), with a localization mechanism distinct from non-Hermitian skin accumulation, which localizes the \emph{entire} bulk continuum.   
We show that scale-sensitive non-Hermitian systems can generically admit a different fate: as we scale up the system size, a detached edge-localized eigenstate can remain sharply visible over a broad window until a critical scale is reached, where it forms an ephemeral bound state in the continuum (BIC) of the open-boundary bulk before being absorbed (entocytosed) at even larger system sizes.
We call this phenomenon edge state endocytosis.  
Its mechanism is fundamentally traced to the Widom expansion of the open-chain characteristic determinant (energy dispersion equation) into contributions corresponding to admissible non-Bloch mode subsets.   
Each subset contribution factorizes into a boundary-projected Green's function (proj-GF) determinant, which encodes lattice truncation, and a subset-resolved bulk propagation factor, which encodes the system size dependence.   
We uncover the fundamental distinction: TDL edge states are zeros of the leading-
subset proj-GF determinant, whereas endocytosed states are a hitherto-ignored class of hidden proj-GF zeros from subleading subsets that control the spectrum at finite sizes. 
Due to its fundamental mathematical origin, the endocytosis mechanism is completely platform-independent, occurring generically without fine-tuning when isolated edge states, topological or otherwise, are subject to non-Hermitian couplings that generate the requisite non-locality. 
Our new framework quantitatively predicts the endocytosis scale and sheds light on how its intricate competitive mechanism can be revealed through experimentally relevant Green's functions. 
Overall, this work identifies novel scale-dependent non-Hermitian physics as a source of robust experimental salient spectral features not captured by the TDL, such as new Green's function responses and flux insertion smoking guns associated with state endocytosis.
\end{abstract}

\date{\today}
\maketitle

\noindent\textcolor{darkgreen}{\textit{Introduction. --}}
An isolated edge state is usually interpreted as a thermodynamic-limit object: once protected, it should survive as a localized state as the system size tends to infinity~\cite{Hatsugai1993chern,Kitaev_2001unpaired,Delplance2011Zak,kane2005quantum,hasan2010colloquium,qi2011topological}.  This expectation is natural in Hermitian band theory, where finite chain spectra are merely discretized versions of already well-defined bulk spectra.  Non-Hermitian systems with broken reciprocity are fundamentally different, however, because their eigenstates are extremely sensitive to boundary conditions, built from complex Bloch factors $\beta=e^{ik}$ which encode system-wide amplification or decay under open boundary conditions (OBCs). Such sensitivity is symptomatic of emergent non-locality, which gives rise to the non-Hermitian skin effect (NHSE) in both non-interacting~\cite{HatanoNelson1996Localization,HatanoNelson1997Vortex,Lee2016nonH,song2019non,yao2018edge,yokomizo2019non,okuma2020quantum,Ashida2020NHPhysics,Bergholtz2021Exceptional,OkumaSato2023Review,Lee2019Hybrid,kawabata2020higher,kunst2018biorthogonal,Yang2026Reversing,Shen2025hyperbolic,kawabata2018anomalous,yao2018non,Leykam2017EdgeModes,MartinezAlvarez2018Robust,Lieu2018SSH,Yin2018Winding,ImuraTakane2019Generalized,Herviou2019SVD,YokomizoMurakami2020PTEP,Zhu2021Delocalization,Borgnia2020Boundary,Zhu2024Review} and interacting many-body systems~\cite{xu2025excitonic,Alsallom2022Fate,Shen2022cluster}.

\begin{figure}[t]
\centering
\maybegraphics[width=\columnwidth]{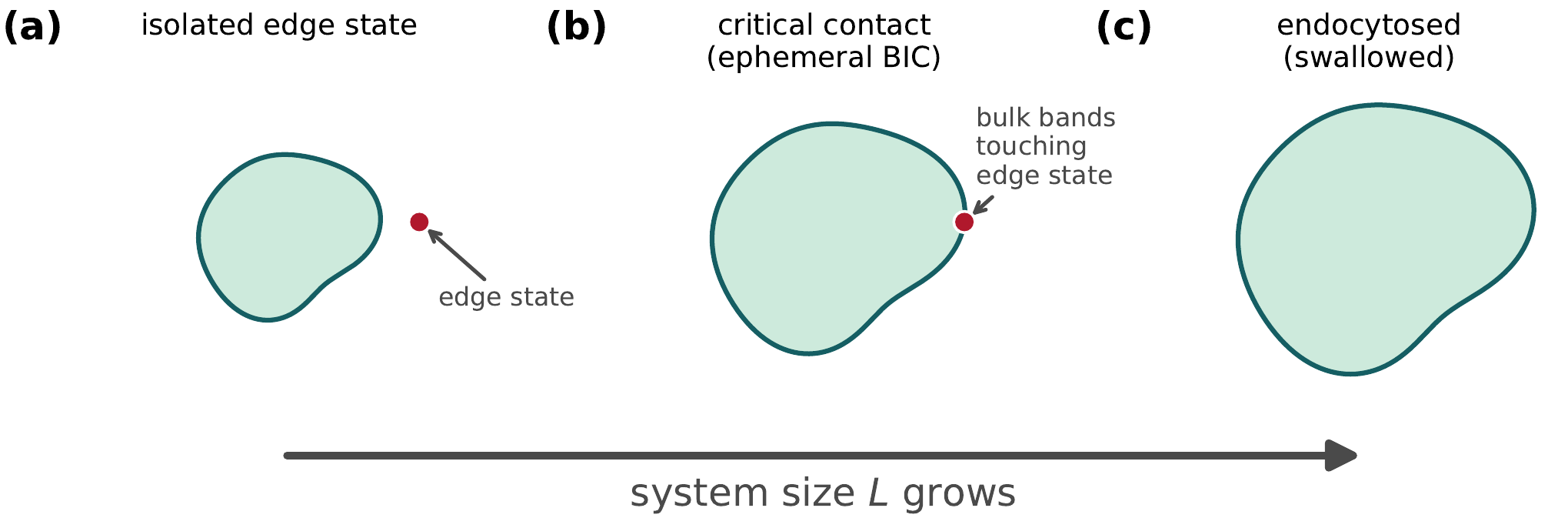}{Schematic showing an isolated edge state, contact with the bulk spectrum, and the endocytosed state as system size grows.}
\caption{\label{fig:main_schematic}
Schematic of edge-state endocytosis in the complex energy plane, which generically occurs when isolated edge modes (topological or otherwise) encounters emergent non-locality stemming from non-Hermitian couplings.    
As the OBC system size $L$ increases, an isolated edge eigenmode (red) gradually encounters the bulk spectrum (green loop) and is swallowed up (endocytosed).  
Unlike in usual NHSE systems, the OBC bulk spectrum encloses finite area in such physical setups with competitive non-locality, mathematically traced to the nontrivial scaling of the subset contributions $S$ in Eq.~\eqref{eq:main_finite_sum}.
	}
\end{figure}

In more sophisticated cases, this non-locality can also be manifested as unconventional scaling behavior, such as in critical NHSE lattices~\cite{li2020critical,yokomizo2021scaling,Islam2022,qin2023universal,Qin2025Many,Liu2022Helical} or systems subject to  generalized boundary conditions~\cite{guo2021exact,Long2025}, disorder~\cite{claes2021skin,zhang2023bulk,Longhi2025Erratic,Jiang2019Interplay} or impurities~\cite{li2021impurity,liu2021exact,Paolo2023Anomalous,cheng2025stochasticity}. 
Yet, edge localization is not always linked to the NHSE -- topological mechanisms,
amid other physical localization mechanisms in non-Hermitian contexts~\cite{lee2022exceptional,zou2024experimental,liu2025non,yang2025beyond}, can also produce robust bound states. It has remained as an open question whether the interplay of such robust non-NHSE states with unconventional non-Hermitian critical scaling can give rise to unprecedented new phenomena.


In this work, we uncover a radically new class of non-Hermitian phenomena which we name ``Edge State Endocytosis". As illustrated in Fig.~\ref{fig:main_schematic}, under OBCs, an isolated edge state or flat band (red) existing at smaller system sizes $L$ gets ``swallowed up" or \emph{endocytosed} by the bulk spectrum (green loop) as $L$ increases. Interestingly, the almost-swallowed state appears as a bound state in the bulk continuum (BIC) briefly at the particular $L$ where critical contact occurs.

This unique scaling-induced bound state destruction is unrelated to disorder broadening, and occurs without fine tuning. As elaborated later, it is a generic phenomenon due to the competition between ``subsets" $S$ of bulk non-Bloch modes that appear in the determinant expansion of the energy dispersion polynomial. While such expansions have been introduced decades ago in classic works on block Toeplitz matrices~\cite{Widom1974BlockToeplitz,Widom1976BlockToeplitzII} by Widom, the intrinsic competition among the expanded terms have never been linked to any saliently measurable physical consequence, not least a new type of scaling phenomenon. Our endocytosis phenomenon is predicted to occur rather ubiquitously, whenever the parent Hamiltonian contains an isolated edge mode (topological or not), accompanied by ``auxiliary" bands that introduce the requisite emergent non-locality.


Theorems~\ref{thm:finite_expansion} and~\ref{thm:pgf_edge} make the distinction between our endocytosis mechanism and ordinary band merging precise. By combining an exact finite-$L$ subset expansion with the projected Green's function (proj-GF)~\cite{Peng2017boundary,Slager2015} edge-state criterion, we elucidate how boundary matching and system size scaling can conspire to make the state swallowing inevitable. The boundary prefactor tests lattice truncation, while the bulk factor decides which subset dominates as $L$ grows.  
A conventional thermodynamic limit (TDL) edge state occurs when the relevant proj-GF zero belongs to the subset that dominates asymptotically.  
Endocytosis occurs when the zero is hidden in a subleading subset that can control finite chains because the asymptotically dominant boundary mismatch is small but nonzero.  At critical contact, this gives an \emph{ephemeral BIC}: a bound state in the OBC non-Bloch continuum at finite $L$.  Unlike in conventional BICs~\cite{Hsu2016BIC,Hsu2013,Sakotic2023}, the bulk embedding is generated by the scaling-driven subset crossover rather than known protection mechanisms inside a fixed continuum.

\begin{figure}[t]
\centering
\maybegraphics[width=\columnwidth]{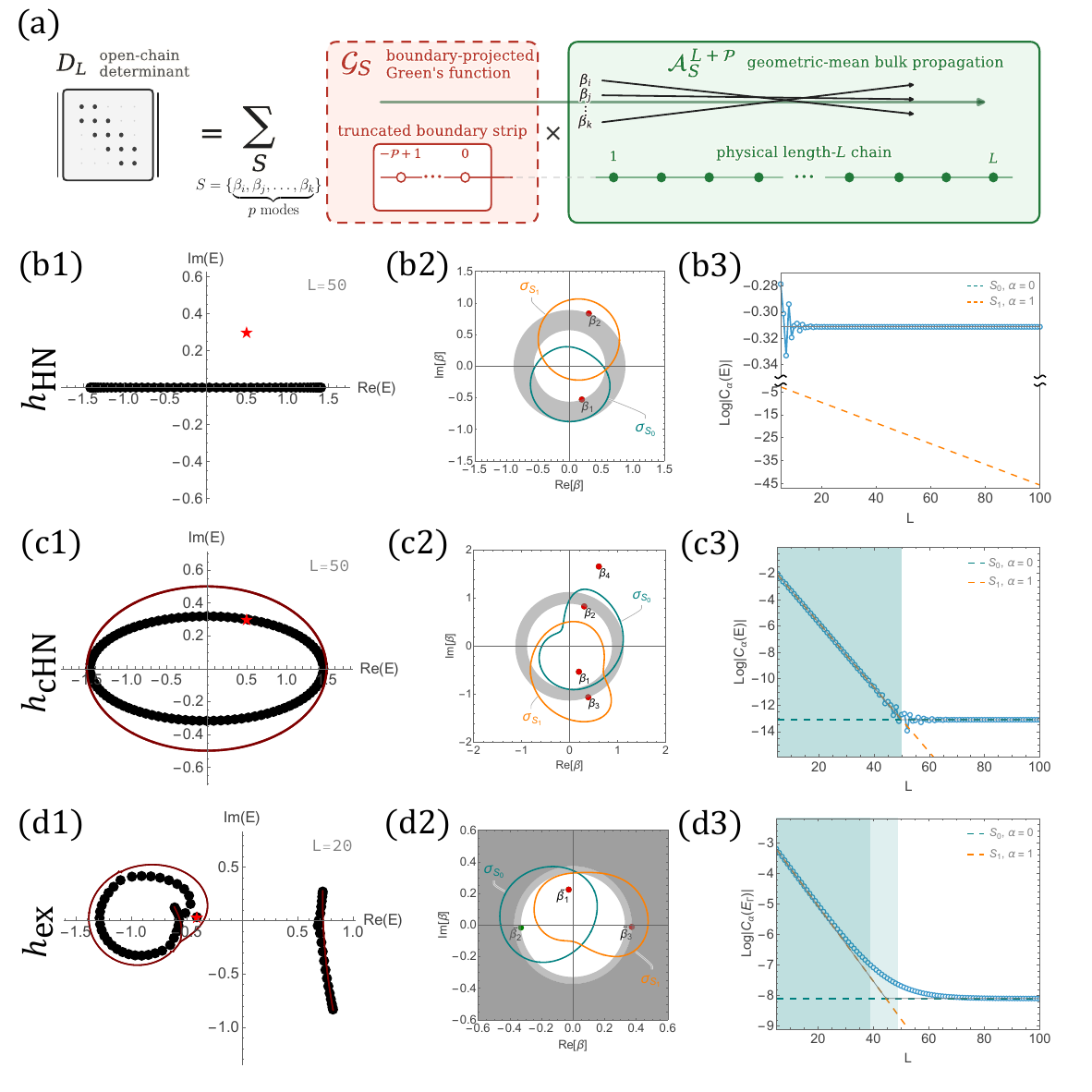}{Subset competition.}
\caption{\label{fig:main_subsets}
Root-subset interpretation of Theorem~\ref{thm:finite_expansion}.  (a) Each subset $S$ selects roots, carries a bulk factor $\calA_S^{L+\PP}$, and has boundary prefactor $\calG_S$.  (b1)--(b3) Single Hatano--Nelson chain $h_{\rm HN}(\beta)=\beta+t/\beta$ with $t=1/2$.  (c1)--(c3) Coupled Hatano--Nelson model $h_{\rm cHN}(\beta)$ with $t=1/2$ and $\Delta=1/1000$.  (d1)--(d3) Main model $h_{\rm ex}$ in Eq.~\eqref{eq:main_example_symbol}.  Panels (b1)--(d1) show finite-$L$ OBC spectra as black dots, and panels (c) and (d) use subset labels $S_0,S_1$.  The five-pointed star in (b1)--(d1) identifies the energy for root panels (b2)--(d2) and competition panels (b3)--(d3).  In (b3)--(d3), dashed curves show $\log\abs{C_\alpha}$ for the competing subset contributions, while blue open circles show the corresponding full finite-chain determinant $D_L$.  Red solid curves in (c1) and (d1) denote the TDL open-boundary bulk spectrum, derived analytically from the GBZ condition.  In (c3) and (d3), cyan shading shows the pre-asymptotic regime where the leading subset loses dominance; in (d3), the lighter region indicates critical contact.  In (d), $E_\Gamma$ samples a loop of radius $r=0.01$ around $E_\ast$, giving $L_c\approx 40$.  Green/orange contours in (b2)--(d2) are representative subset contours of Theorem~\ref{thm:finite_expansion}, and the gray annuli denote GBZ modulus gap $\abs{\beta_p}<\abs{\beta}<\abs{\beta_{p+1}}$.  In (d2), red roots come from the parent band and the green root from the auxiliary band.  Darker and lighter gray annuli mark the GBZ modulus gaps of the parent Hamiltonian and the full Hamiltonian $h_{\rm ex}$, respectively.
}
\end{figure}

\noindent\textcolor{darkgreen}{\textit{Mechanism setup: Bulk Toeplitz subset expansion. --}}
An $m$-component one-dimensional lattice under OBC with finite left and right hopping ranges of $\PP$ and $\QQ$ unit cells is represented in real space by a finite block Toeplitz matrix~\cite{SchmidtSpitzer1960Toeplitz,BottcherGrudsky2005Banded,Bottcher2000Toeplitz,Tilli1998}, and its corresponding Bloch Hamiltonian is
\begin{equation}
 h(\beta)=\sum_{r=-\PP}^{\QQ}h_r\beta^r.
\label{eq:main_symbol}
\end{equation}
We denote by $H_L$ the length-$L$ open-chain truncation generated by the same hopping matrices $h_r$ (i.e. the OBC version of $h(\beta)$).  To keep track of the left and right hopping ranges in terms of the physical sites, we write $p=m\PP$ and $q=m\QQ$.  We first assume the left-most and right-most hopping matrices are non-singular, $\det h_{-\PP}\det h_{\QQ}\neq 0$, for mathematical brevity.  At a complex energy $E$, the non-Bloch determinant is
\begin{equation}
 D(\beta,E)\equiv\det(E-h(\beta)).
\label{eq:main_characteristic_equation}
\end{equation}
With nonsingular extreme hoppings, $D(\beta_i,E)=0$ gives $p+q$ finite non-Bloch factors ordered by $\abs{\beta_1}\le\abs{\beta_2}\le\ldots\le\abs{\beta_{p+q}}$, with right null vectors $u(\beta_i)$ satisfying $(E-h(\beta_i))u(\beta_i)=0$.  An {OBC eigenstate} solution is a superposition $\sum_i c_i\beta_i^x u(\beta_i)$, and lattice truncation selects allowed superpositions.  For the expansion below, a \emph{subset} $S$ denotes a local $p$-root choice, with $p$ equal to the left-boundary strip dimension, typically $S=\{\beta_{i_1},\ldots,\beta_{i_p}\}$.  For $p=q=2$, examples include $\{\beta_1,\beta_2\}$ and $\{\beta_1,\beta_3\}$.

At finite $L$, bulk transfer and boundary matching are separate conditions to be satisfied by the OBCs.
A finite chain asks both how a selected root set scales through the chain and whether it can cancel the boundary mismatch.
To keep track of the OBC eigenenergies, we focus on the zeroes of the finite-chain energy dispersion determinant 
\begin{equation}
D_L(E)\coloneqq\det(E-H_L).
\end{equation}
A Widom-type determinant expansion~\cite{Widom1974BlockToeplitz,Widom1976BlockToeplitzII,BottcherGrudsky2005Banded,BottcherSilbermann2006Toeplitz,TracyWidom2002ToeplitzMinors,BumpDiaconis2002ToeplitzMinors} gives the following decomposition:

\begin{theorem}[subset expansion]
\label{thm:finite_expansion}
On a collision-free local root domain, choose a contour $\sigma_S$ that is positively oriented, winds once around the origin, and encloses only the roots in $S$ but no others.  Define $\GG_S(x,y,E)\coloneqq(2\pi i)^{-1}\oint_{\sigma_S}\beta^{x-y}(E-h(\beta))^{-1}\dd\beta/\beta$, and let $P_\partial$ project onto this boundary block.  For every integer $L>\max(\PP,\QQ)$,
\begin{equation}
\begin{aligned}
D_L(E)&=\sum_S \calA_S(E)^{L+\PP}\calG_S(E),\\
\calA_S(E)
&\coloneqq\exp\left\{\frac{1}{2\pi i}\oint_{\sigma_S}
\log\det\bigl(E-h(\beta)\bigr)\frac{\dd\beta}{\beta}\right\},\\
\calG_S(E)
&\coloneqq\det\left(
\frac{1}{2\pi i}\oint_{\sigma_S}\beta^{\mu-\nu}
\bigl(E-h(\beta)\bigr)^{-1}\frac{\dd\beta}{\beta}
\right)_{\mu,\nu=0}^{\PP-1}\\
&=\det(P_\partial\GG_S(E)P_\partial).
\end{aligned}
\label{eq:main_finite_sum}
\end{equation}
The sum over all admissible $S$ gives an exact finite-$L$ subset decomposition of $D_L$.
\end{theorem}
The finite-chain OBC spectrum satisfies $D_L(E)=0$.  The determinant in $\calG_S$ is over the $p$-dimensional boundary block.  The first integral gives the geometric mean bulk factor $\calA_S$, while the last equality displays $\calG_S$ as the boundary-projected Green's-function determinant.  Thus $\calA_S^{L+\PP}$ carries the length dependence, while $\calG_S$ tests boundary compatibility.  This separation is illustrated in Fig.~\ref{fig:main_subsets}, and organizes the distinction below between ordinary bulk competition and hidden boundary-prefactor zeros.  The proof is given in the Supplemental Material, Secs.~SI and SII~\cite{supplemental_material}.

Before deriving what gives rise to the endocytosis mechanism, we briefly discuss two pedagogical examples. The simplest example, the Hatano--Nelson model, illustrates how to read the two factors in the Widom expansion: $\calA_S$ is the bulk exponential for a selected root set, while $\calG_S$ is its boundary-compatibility amplitude.  For a single Hatano--Nelson chain $h_{\rm HN}(\beta)=\beta+t/\beta$, the simple roots obey $\beta_1\beta_2=t$, and the two terms have $\calA_{\{1\}}=t/\beta_1=\beta_2$ and $\calA_{\{2\}}=t/\beta_2=\beta_1$, with residue prefactors $\calG_{\{1\}}=1/(\beta_2-\beta_1)$ and $\calG_{\{2\}}=1/(\beta_1-\beta_2)$.  The OBC continuum is the bulk competition $|\calA_{\{1\}}|=|\calA_{\{2\}}|$, equivalently $|\beta_1|=|\beta_2|$.

As the next pedagogical example, the coupled Hatano--Nelson model in Fig.~\ref{fig:main_subsets}(c1)--(c3), $h_{\rm cHN}(\beta)=\left(\begin{smallmatrix}\beta+t/\beta&\Delta\\ \Delta&t\beta+1/\beta\end{smallmatrix}\right)$, shows the same separation for two-root subsets.  If $\beta_i$ are the four roots of $\det(E-h_{\rm cHN}(\beta))=0$, then $S=\{i, j\}$ has $\calA_{\{i, j\}}=t/(\beta_i\beta_j)$ and $\calG_{\{i, j\}}=(\beta_j-\beta_i)^2(t-\beta_i\beta_j)(t-1/(\beta_i\beta_j))/\prod_{\ell=i,j}[4t\beta_\ell^3-3E(1+t)\beta_\ell^2+2(E^2+1+t^2-\Delta^2)\beta_\ell-E(1+t)]$.  Here $\calA_{\{i, j\}}$ ranks subsets by their bulk growth or decay across the chain, while $\calG_{\{i, j\}}$ tests whether the two selected modes actually span the two boundary degrees of freedom.  In the weak-coupling regime ($\Delta\ll1$), the leading subset $S_0$ can have $\calG_{S_0}=\mathcal O(\Delta^2)$, so it is exponentially favored by $\calA_{S_0}^{L+\PP}$ but parametrically suppressed by boundary matching.  This is the conceptual separation between bulk scaling and boundary compatibility.

\begin{figure*}[t]
\centering
\maybegraphics[width=0.9\textwidth]{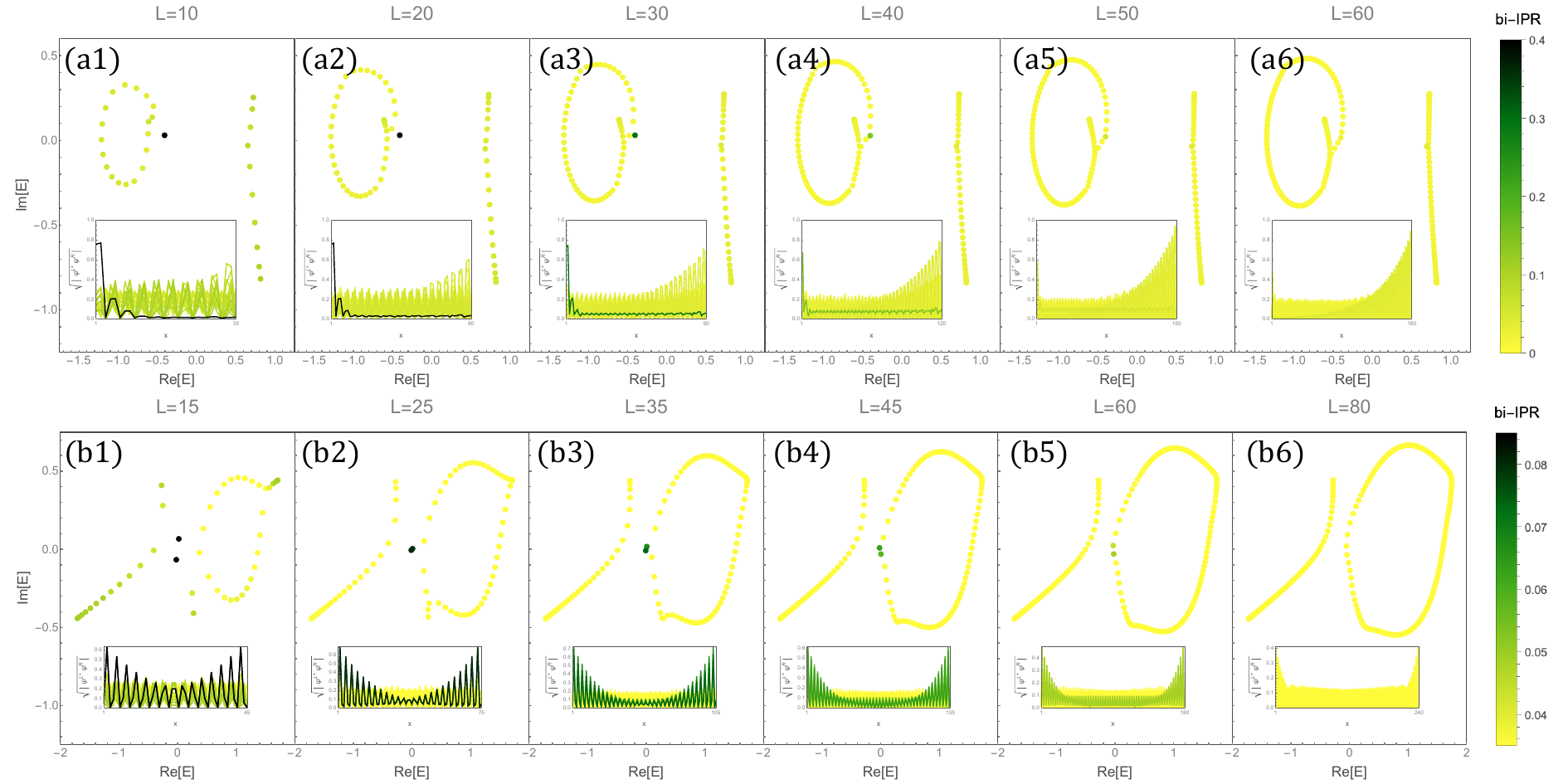}{OBC spectra at finite L for the $h_{\rm ex}$ three-band model, colored by biorthogonal inverse participation ratio.}
\caption{\label{fig:main_biIPR}
The endocytosis process: OBC spectra at a sequence of system sizes $L$, colored by bi-IPR.  Panel (a) uses model $h_{\rm ex}$ in Eq.~\eqref{eq:main_example_symbol}.  Panel (b) uses a separate demonstration model $h_{\rm ex2}(\beta)=\left(\begin{smallmatrix}0&\beta+1.2+0.40i&1.0\times10^{-3}\\ \beta^{-1}+0.40+0.40i&0&1.0\times10^{-3}\\ 1.0\times10^{-3}&1.0\times10^{-3}&1+0.91\beta\end{smallmatrix}\right)$, which differs from $h_{\rm ex}$ by starting from a non-Hermitian SSH parent model with nontrivial band topology.  With increasing system size, the isolated edge eigenenergy loses localization and is absorbed by the OBC bulk spectrum at finite $L$. Inset: biorthogonal wavefunction density.
}
\end{figure*}

Equation~\eqref{eq:main_finite_sum} expresses the finite determinant in a complex-root subset basis, where finite-chain eigenenergies arise from cancellations among terms with length-dependent amplitudes and phases.

\noindent\textcolor{darkgreen}{\textit{Usual spectra and the endocytosis case. --}}
The distinction between bulk bands, ordinary edge states, and endocytosis is set by which subset terms control Eq.~\eqref{eq:main_finite_sum}.
Here ``bulk'' means the continuum-like OBC spectral set, even when its eigenvectors are skin-localized.
Away from a bulk-band degeneracy, subset labels are ordered by decreasing $|\calA_S|$: $S_0,S_1,S_2,\ldots$.
\begin{enumerate}[label=\Roman*., leftmargin=1.8em, itemsep=1pt, topsep=2pt, parsep=0pt]
\item OBC bulk bands.  In the TDL these are the loci of the OBC bulk spectrum, equivalently fulfilled GBZ condition~\cite{yao2018edge,SchmidtSpitzer1960Toeplitz,yokomizo2019non,YokomizoMurakami2020PTEP,Yang2020AuxGBZ}, where two asymptotically dominant subsets have comparable bulk factors, for example $|\calA_{S_a}|=|\calA_{S_b}|$.  At finite $L$, OBC eigenenergies are still zeros of the full determinant $D_L(E)$, and these zeros come from cancellations among subset terms.  In the two-subset competition sketched in Fig.~\ref{fig:main_subsets}(c1)--(c3), the relavent subsets are $S_0$ and $S_1$, and the cancellation that predicts at which $L$ bulk spectrum sweeps over $E$ is between the full contributions $\calA_{S_0}^{L+\PP}\calG_{S_0}$ and $\calA_{S_1}^{L+\PP}\calG_{S_1}$, not only between their bulk factors.  Additional subset terms may also be present.
\item TDL edge states.  These occur away from the GBZ condition, where the large-$L$ determinant is controlled by a leading subset $S_0$ and an edge state requires $\calG_{S_0}(E_\ast)=0$.  This is because all subleading terms are exponentially suppressed in Eq.~\eqref{eq:main_finite_sum}, leaving the leading boundary prefactor to set the local zero.
\item Edge state endocytosis.  This occurs when at least one subleading subset carries a hidden boundary-prefactor zero absent from the leading subset, so finite $L$ can show a sharply localized eigenvalue before the leading hierarchy takes over.  This happens because the leading prefactor is small but nonzero, allowing a subleading subset term to overtake it over a finite length window.
\end{enumerate}

For the local competition, choose a loop away from the TDL bulk-band condition.  Then $S_0$ is asymptotically leading as $L$ grows.  The simplest nontrivial two-subset case, realized by $h_{\rm ex}$ in Eq.~\eqref{eq:main_example_symbol}, has $S_1$ carrying the hidden zero.  Dividing $D_L$ by $\calA_{S_0}^{L+\PP}$ gives
\begin{equation}
\frac{D_L(E)}{\calA_{S_0}(E)^{L+\PP}}
=\underbrace{\calG_{S_0}(E)}_{C_0(E)}
+\underbrace{\eta_{10}(E)^{L+\PP}\calG_{S_1}(E)}_{C_1(E,L)}
+\mathcal O(\eta_{20}^{L+\PP}),
\label{eq:main_two_subset}
\end{equation}
where $\eta_{ij}(E)=\calA_{S_i}(E)/\calA_{S_j}(E)$ is the relative bulk factor, locally the ratio of selected root products~\cite{supplemental_material}.  The $\mathcal O(\eta_{20}^{L+\PP})$ term is set by the largest omitted subset $S_2$.  After the leading factor is removed, $\eta_{i0}$ determines how rapidly subset $S_i$ is suppressed with length.
An eigenenergy is a zero of the left-hand side, so Eq.~\eqref{eq:main_two_subset} is the local competition criterion: the leading-subset boundary term $C_0(E)$ must be canceled by the length-dependent subleading contribution $C_1(E,L)$, up to the omitted smaller terms.
Given that $|\eta_{i0}|<1$ for $i\ge1$ and $\calG_{S_0}$ is not anomalously small, the powers $\eta_{i0}^{L+\PP}$ make all subleading terms exponentially irrelevant and local zeros are governed by $\calG_{S_0}$.  
If $\calG_{S_0}(E_\ast)=0$, Eq.~\eqref{eq:main_two_subset} leaves only exponentially small residual terms, so a zero of $D_L$ converges to $E_\ast$ and gives an ordinary TDL edge state.

Endocytosis is the complementary small-prefactor case: $S_0$ has the largest bulk factor and a small but nonzero $\calG_{S_0}(E_\ast)$, while $S_1$ has the nearby hidden zero $\calG_{S_1}(E_\ast)=0$.  At small and intermediate lengths, $C_1$ can overtake the leading residual and create a finite-chain eigenvalue near $E_\ast$.  At large lengths, $|\eta_{10}|^{L+\PP}$ suppresses $C_1$, so the hidden zero remains subset-resolved but no longer gives a zero spectrally separated from the OBC bulk spectrum at that size.

The crossover length follows from the balance of $C_1$ with $C_0$.  With $\langle\cdots\rangle_\Gamma$ denoting the average over sampled points on a small loop $\Gamma$ around $E_\ast$,
\begin{equation}
\begin{aligned}
L_c+\PP
&\sim
\frac{\left\langle\log\!\left|\calG_{S_1}(E)/\calG_{S_0}(E)\right|\right\rangle_\Gamma}
{-\left\langle\log|\eta_{10}(E)|\right\rangle_\Gamma}
\\
&=
\frac{\left\langle\log\!\left|\calG_{S_1}(E)/\calG_{S_0}(E)\right|\right\rangle_\Gamma}
{\left\langle\log\!\left|\calA_{S_0}(E)/\calA_{S_1}(E)\right|\right\rangle_\Gamma}.
\end{aligned}
\label{eq:main_crossover_length}
\end{equation}
This estimate sets the pre-asymptotic window over which the hidden boundary-prefactor zero can be visible in the full spectrum.  When the OBC bulk spectrum at that size enters the same neighborhood, the loss of a separated zero is observed as absorption into the OBC bulk spectrum.
Figure~\ref{fig:main_subsets}(d3) shows this balance explicitly: as the system size $L$ increases, the sampled log-amplitude of $C_1(E,L)$ decreases with $\langle\log\abs{\eta_{10}(E)}\rangle_\Gamma<0$ and 
crosses that of $C_0(E)$ at $L_c\approx 40$, where determinant control switches between the two subsets.  Dominance estimates, eigenenergy displacement, 
and the precise endocytosis interval are given in the Supplemental Material, Sec.~SIII~\cite{supplemental_material}.

\noindent\textcolor{darkgreen}{\textit{Examples and numerical signatures. --}}
The numerical example used here is a three-band nearest-neighbor chain with Bloch Hamiltonian
\begin{equation}
h_{\rm ex}(\beta)=h_{-1}\beta^{-1}+h_0+h_1\beta,
\label{eq:main_example_symbol}
\end{equation}
with $\operatorname{rank}h_{-1}=2$ and
\begin{equation}
\begin{aligned}
h_{-1}&=\left(\begin{smallmatrix}
0.26+0.20i&0.13+0.19i&0\\
-0.43+0.20i&-0.43-0.015i&0\\
0&0&0
\end{smallmatrix}\right),\\[-1pt]
h_0&=\left(\begin{smallmatrix}
-0.090+0.16i&0.34+0.35i&0.010\\
0.28-0.42i&0.11-0.37i&0\\
0.010&0&-1
\end{smallmatrix}\right),\\[-1pt]
h_1&=\left(\begin{smallmatrix}
0.37-0.065i&-0.22-0.040i&0\\
-0.045-0.42i&0.0050+0.30i&0\\
0&0&-1.8
\end{smallmatrix}\right).
\end{aligned}
\label{eq:main_example_hoppings}
\end{equation}
The target is $E_\ast\simeq -0.4079+0.0302i$, with root ordering detailed in the Supplemental Material, Sec.~SIV~\cite{supplemental_material}.  For singular extreme hoppings, the same identity uses the pole count $\wtp=-\ord_{\beta=0}\det(E-h(\beta))$ as the finite-root count in each finite-root subset, while the boundary factor remains the $p\times p$ determinant $\det(P_\partial\GG_S P_\partial)$.  Details are in the Supplemental Material, Secs.~SI and SII~\cite{supplemental_material}.  Here $p=3$ and $\wtp=2$, with the remaining boundary column supplied by the local origin contribution in $P_\partial\GG_S P_\partial$.  The auxiliary root changes the subset ordering, so the inherited parent edge zero becomes a hidden boundary-prefactor zero of a subleading subset rather than a TDL edge mode.

Ordering those finite roots by modulus, because the left boundary has effective rank $\tilde p=2$, the two relevant subsets are
\begin{equation}
S_0=\{\tilde\beta_1,\tilde\beta_2\},
\qquad
S_1=\{\tilde\beta_1,\tilde\beta_3\},
\qquad
\eta_{10}=\frac{\calA_{S_1}}{\calA_{S_0}}
=\frac{\tilde\beta_2}{\tilde\beta_3}.
\label{eq:main_rank2_subsets}
\end{equation}
The two subsets differ only by $\tilde\beta_2\leftrightarrow\tilde\beta_3$, as shown in Fig.~\ref{fig:main_subsets}(d1)--(d3): root moduli select $S_0$ in the TDL, while the boundary prefactor can still favor $S_1$ over a finite length window.

Fig.~\ref{fig:main_biIPR} shows OBC spectra at selected lengths colored by the biorthogonal inverse-participation ratio.  Panel (a) uses Eq.~\eqref{eq:main_example_symbol}, while panel (b) is a supplementary demonstration and is not used elsewhere.  The bi-IPR is defined as
\begin{equation}
{\rm bi\text{-}IPR}_{n}
=
\frac{\sum_x \rho_x^2}{\big(\sum_x \rho_x\big)^2},
\label{eq:main_biipr}
\end{equation}
with $\rho_x=\abs{\langle\psi^L_{n,{\rm OBC}}|\hat{x}|\psi^R_{n,{\rm OBC}}\rangle}$ and $\hat{x}$ projecting onto unit cell $x$, including all sublattices.  Larger bi-IPR means stronger localization.  In Fig.~\ref{fig:main_biIPR}(a), the circled eigenenergy is detached from the OBC bulk spectrum at that length and the inset shows strong boundary localization.  As $L$ grows, the subset competition in Eq.~\eqref{eq:main_two_subset} expands the OBC bulk spectrum at finite $L$ toward the circled eigenenergy.  Figure~\ref{fig:main_biIPR}(b) shows the post-contact stage in a supplementary model, where the isolated edge eigenenergy is no longer resolved.

\begin{figure}
\centering
\maybegraphics[width=\columnwidth,keepaspectratio]{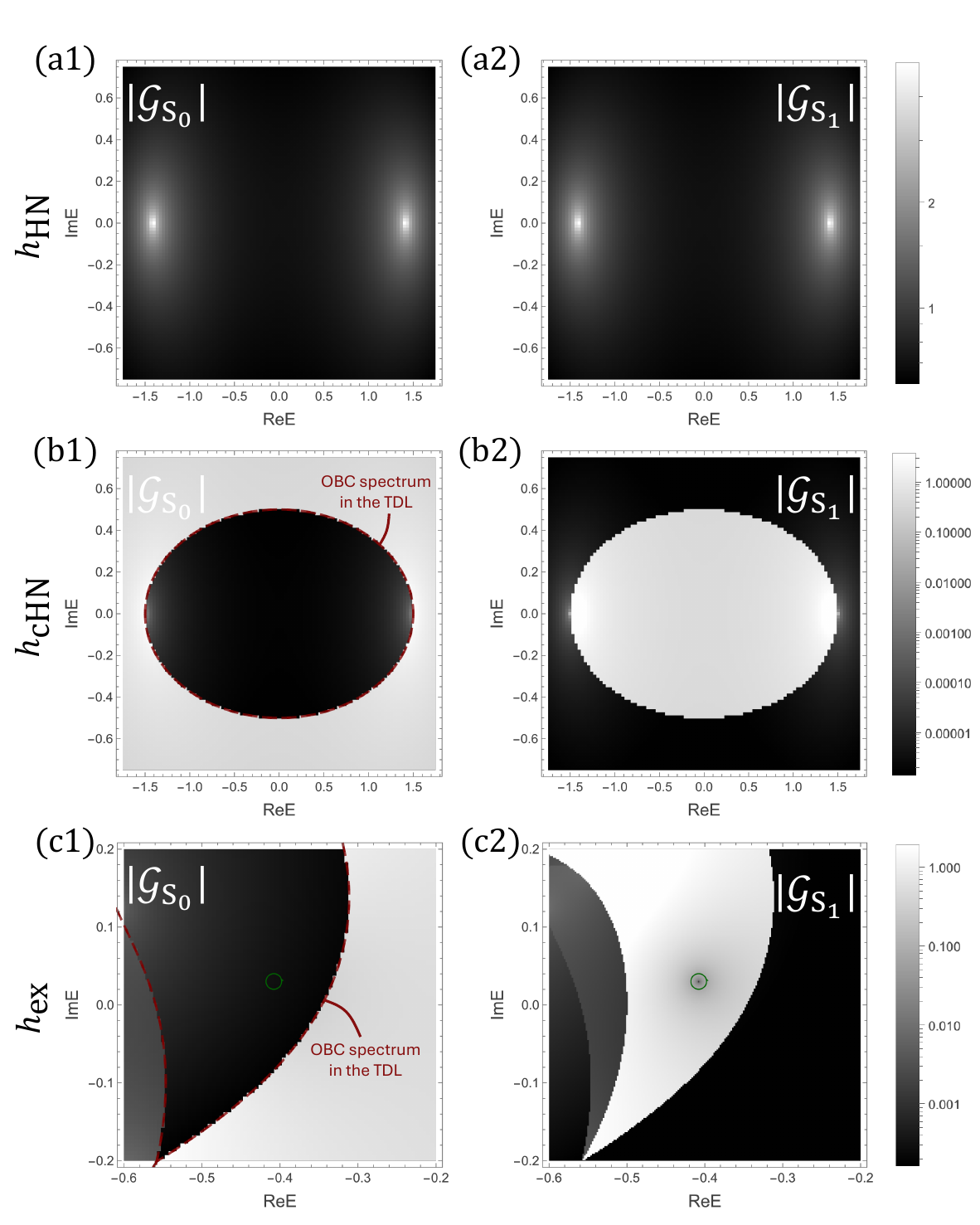}{Subset-resolved boundary-prefactor diagnostics for Hatano--Nelson, coupled Hatano--Nelson, and the main model.}
\caption{\label{fig:main_winding}
Uncovering the anatomy of subset-resolved boundary-prefactors through Green's functions.  (a1)--(a2) $\abs{\calG_{S_0}(E)}$ and $\abs{\calG_{S_1}(E)}$ for the two one-root subsets of the single-chain Hatano--Nelson model $h_{\rm HN}$ with $t=1/2$, serving as basic examples.  (b1)--(b2) Corresponding diagnostic for representative two-root subsets of the coupled Hatano--Nelson model $h_{\rm cHN}$ with $t=1/2$ and $\Delta=1/1000$.  (c1)--(c2) Main-text model $h_{\rm ex}$ in Eq.~\eqref{eq:main_example_symbol}, where $\nu_i\equiv\nu_{S_i}(\Gamma)$ and the green loop $\Gamma$ in (c1) and (c2) gives the nontrivial relative winding $\Delta\nu_{10}\equiv\nu_1-\nu_0=1$.  Dashed red curves in (b1) and (c1) denote the OBC bulk spectrum in the TDL.  The zeros diagnose the TDL fate of edge states, as detailed in Table~\ref{tab:subset_winding}.
}
\end{figure}

\noindent\textcolor{darkgreen}{\textit{Green's function diagnostics. --}}
The boundary prefactor $\calG_S$ can be evaluated without diagonalizing large finite chains.  The subset contour Green's function $\GG_S$ is restricted to the non-Bloch modes selected by $S$, following standard non-Bloch resolvent constructions~\cite{xue2021simple,fu2023anatomy}, and its projection to the boundary strip gives the finite-dimensional lattice-truncation test.  Green's-function formulations of boundary and topological diagnostics are standard~\cite{EssinGurarie2011Green,Slager2015,Alase2017,Zirnstein2021Green,Borgnia2022,HuWang2023}.

\begin{theorem}[proj-GF edge-state criterion]
\label{thm:pgf_edge}
For an admissible subset $S$ and a regular energy $E_\ast$, $\calG_S(E_\ast)=0$ if and only if the projected boundary block $P_\partial\GG_S(E_\ast)P_\partial$ is singular.  Equivalently, the non-Bloch modes selected by $S$ can be linearly combined so that the lattice-truncation residual vanishes, giving an edge state solution satisfying the OBC boundary condition.
The proof is given in the Supplemental Material, Sec.~SII~\cite{supplemental_material}.
\end{theorem}
We call these projected Green's-function determinant zeros proj-GF zeros.  When the same subset controls the finite determinant in Eq.~\eqref{eq:main_finite_sum}, the proj-GF zero $\calG_S(E_\ast)=0$ produces an edge-localized eigenvalue.

This interpretation makes the zero-count test local in the complex-energy plane: it can be evaluated on loops where the chosen subset remains analytic.  For a positively oriented loop $\Gamma$ around a candidate endocytosis point $E_\ast$, define
\begin{equation}
\nu_S(\Gamma)
\equiv
\frac{1}{2\pi i}\oint_\Gamma
\dd\log\calG_S(E).
\label{eq:main_subset_winding}
\end{equation}
This count distinguishes the TDL fate of nearby edge states: leading proj-GF zero, hidden subleading zero, or trivial.  It is evaluated on loops without proj-GF poles or after removing common analytic factors, and is distinct from point-gap or spectral windings for band topology and the NHSE~\cite{shen2018topological,gong2018topological,kawabata2019symmetry,zhang2019correspondence,okuma2020quantum}.  For a subleading subset $S_a$ that competes over a length interval, define $\Delta\nu_{a0}\equiv\nu_{S_a}-\nu_{S_0}$.


\begin{table}[tbp]
\caption{\label{tab:subset_winding}
Identifying fundamentally different avenues of OBC state formation from local windings: Subset-winding diagnostic on a pole-free local loop $\Gamma$ around $E_\ast$, away from the TDL bulk-band condition and with $S_0$ the unique leading subset.
}
\footnotesize
\centering
\setlength{\tabcolsep}{3pt}
\begin{ruledtabular}
\begin{tabular}{p{0.40\columnwidth}p{0.51\columnwidth}}
Local winding data & Type of OBC state \\
\hline
$\nu_{S_0}\ne0$
&
Leading-subset proj-GF zero; the edge state survives the TDL.
\\
$\nu_{S_0}=0,\ \Delta\nu_{a0}\ne0$
&
Hidden proj-GF zero in $S_a$; if $S_a$ dominates over a length interval, the isolated eigenenergy undergoes endocytosis.
\\
$\nu_{S_0}=0,\ \Delta\nu_{a0}=0$
&
No leading or relative hidden proj-GF zero; locally trivial.
\end{tabular}
\end{ruledtabular}
\end{table}

For $h_{\rm ex}$, $S_a=S_1$ and the path $\Gamma$ in Fig.~\ref{fig:main_winding}(c1)--(c2) gives $\Delta\nu_{10}=1$ after common analytic factors are removed, so $S_1$ carries one local proj-GF zero absent from $S_0$.  Its visibility in the full determinant is then set by the balance of $C_0$ and $C_1$ in Eq.~\eqref{eq:main_two_subset}: nonzero relative winding, a small leading residual, and length-controlled transfer of determinant dominance.

\begin{figure}
\centering
\maybegraphics[width=\columnwidth,keepaspectratio]{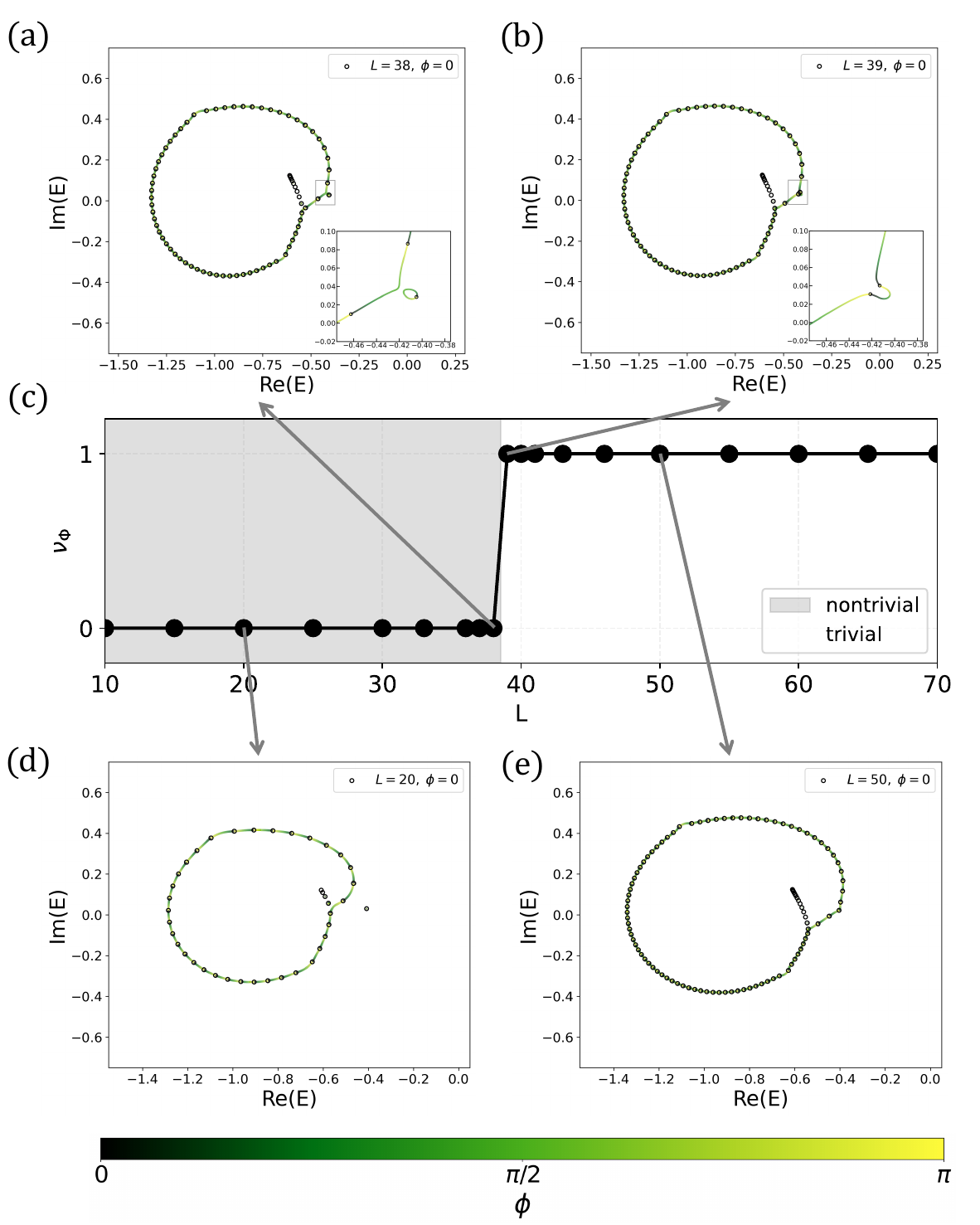}{Weak-coupling-flux finite-chain spectral flow showing the isolated eigenenergy moving toward and into the open-boundary bulk spectrum.}
\caption{\label{fig:main_flux}
Length-dependent flux response of endocytosis model $h_{\rm ex}$ in Eq.~\eqref{eq:main_example_symbol}.  (a)--(b) Critical cases at $L=38,39$, where the isolated eigenenergy is absorbed by the OBC bulk spectrum at those lengths.  (c) The flux winding $\nu_\Phi(L)$ defined in Eq.~\eqref{eq:main_flux_winding} is computed with the small regulator $\xi=0.006$.  (d) At $L=20$ the edge state is fully isolated; (e) at $L=50$ it is fully endocytosed.  As the effective flux is $e^{2i\Phi}$ per plaquette, $0\le\Phi\le\pi$ is one full pumping cycle.
}
\end{figure}

\noindent\textcolor{darkgreen}{\textit{Diagnosing state endocytosis from flux threading. --}}
Gauge flux threading provides a complementary spectral-flow check on the onset of edge state endocytosis~\cite{Laughlin1981,Thouless1983,Lee2020unraveling,Rui2023}.  We use a diagnostic weak-coupling phase twist: the same phase $e^{i\Phi}$ is applied only to the weak parent--auxiliary coupling, 
such as the $(1,3)$ and $(3,1)$ entries of $h_0$ in our three-band realization, rather than to every inter-chain hopping.  Since the same phase is carried by both weak-coupling directions, the pumping flux per plaquette is $e^{2i\Phi}$.  Equivalently, $H_L(\Phi+\pi)$ is related to $H_L(\Phi)$ by a simple gauge transformation, so $0\le\Phi\le\pi$ already gives one full pumping cycle.  Over one such cycle, the finite-chain eigenvalues trace closed loops 
in the complex-$E$ plane.  For the winding plot in Fig.~\ref{fig:main_flux}(c), we fix a reference point $E_{\rm ref}=E_\ast+i\xi$, with small regulator $\xi$, off the flux spectrum and define
\begin{equation}
\nu_\Phi(L)
=\frac{1}{2\pi i}\int_0^\pi \dd\Phi\,\partial_\Phi
\log\det\!\bigl[E_{\rm ref}-H_L(\Phi)\bigr].
\label{eq:main_flux_winding}
\end{equation}
The integer $\nu_\Phi$ is a homotopy-invariant winding of the flux loop at fixed $L$ in $\C\setminus\{E_{\rm ref}\}$, changing only when the loop crosses $E_{\rm ref}$.  Figure~\ref{fig:main_flux} uses this gauge response to track whether the finite-$L$ loop tied to the hidden proj-GF zero has nonzero winding under the weak parent--auxiliary pump, giving a conceptual contrast with nearby continuum states at critical contact.
The detailed flux convention and numerical implementation are given in the Supplemental Material, Sec.~SVIII~\cite{supplemental_material}.

\noindent\textcolor{darkgreen}{\textit{Generic construction. --}}
A generic route starts from a parent boundary mode and does not rely on its topological origin~\cite{MongShivamoggi2011Dirac,Delplace2011Zak,Zhong2024TopologicalNature,Song2019RealSpace,Li2021ScaleFreeLocalization}.  The parent supplies the proj-GF zero that becomes hidden after auxiliary-root insertion, weak auxiliary coupling makes the neighboring leading-subset residual small but nonzero, and the inserted root tunes the bulk-factor ratio.  A two-crossover example from a non-topological parent edge mode is given in the Supplemental Material, Sec.~SVI~\cite{supplemental_material}.

At the Bloch Hamiltonian level, this means a weak block coupling of the form
\begin{equation}
h(\beta)=
\begin{pmatrix}
h_{\rm parent}(\beta)&\kappa T_{\rm pa}(\beta)\\
\kappa T_{\rm ap}(\beta)&h_{\rm aux}(\beta)
\end{pmatrix},
\qquad 0<|\kappa|\ll1,
\label{eq:main_parent_aux_coupling}
\end{equation}
where $T_{\rm pa}$ and $T_{\rm ap}$ are finite Laurent-polynomial hopping blocks.  Because $0<|\kappa|\ll1$, the parent proj-GF zero is inherited with only perturbative modification, and the auxiliary root drifts only perturbatively.

For a parent edge mode near $E_\ast$, order the relevant roots by modulus and let $S_{\rm parent}=\{\beta_1,\ldots,\beta_{p_p}\}$.  Insert an auxiliary band with one root $|\beta_{p_p-1}|<|\beta_{\rm a}|<|\beta_{p_p}|$.  Then $S_0^{\rm new}=\{\beta_1,\ldots,\beta_{p_p-1},\beta_{\rm a}\}$ is leading, while $S_1^{\rm new}=\{\beta_1,\ldots,\beta_{p_p}\}$ becomes nearby subleading.  Figure~\ref{fig:main_subsets}(d2) realizes this rule: the green root replaces the corresponding parent root in $S_0$.

Thus $S_1^{\rm new}$ inherits the parent proj-GF zero.  Dropping the superscript, weak coupling lifts the leading-subset boundary determinant to $\calG_{S_0}^{(\kappa)}=\mathcal{O}(\kappa^\ell)$, where $\ell$ is the first nonzero order needed to lift the decoupled boundary rank deficiency~\cite{supplemental_material}.  Equation~\eqref{eq:main_two_subset} then applies with $|\calA_{S_0}|>|\calA_{S_1}|$: the hidden proj-GF zero can produce a finite-$L$ edge eigenvalue, but not a spectrally isolated large-$L$ eigenenergy.

A mathematically convenient way to implement this insertion is a one-directional auxiliary orbital,
\begin{equation}
 h_{\rm aux}(\beta)=a-b\beta,
 \qquad \beta_{\rm a}(E)=\frac{a-E}{b},
\label{eq:main_aux_band}
\end{equation}
where $a$ and $b$ set the position of the auxiliary root.  Choosing $(a-E_\ast)/b$ between $\beta_{p-1}$ and $\beta_p$ in modulus makes $S_0^{\rm new}$ exchange one parent root for $\beta_{\rm a}$, while $S_1^{\rm new}$ retains the parent proj-GF zero.  The one-directional form is not essential.  It simply supplies one controllable finite root without introducing extra nearby roots that would obscure the subset exchange.  More general auxiliary bands are allowed when changes in selected-root cardinality are compensated by roots common to the competing subsets, as detailed in the Supplemental Material, Sec.~SV~\cite{supplemental_material}.

\noindent\textcolor{darkgreen}{\textit{Conclusion. --}}
We have shown that an edge-localized eigenenergy in a finite non-Hermitian open chain need not be the precursor of a thermodynamic edge state.  The finite-$L$ expansion of $D_L$ separates boundary matching, encoded in $\calG_S$, from the subset bulk factor $\calA_S$.  A thermodynamic edge state requires a zero of the leading-subset prefactor, whereas endocytosis appears when a hidden proj-GF zero of a subleading subset controls the finite determinant before the leading hierarchy takes over.

The isolated eigenenergy is absorbed when the determinant crosses over from hidden proj-GF-zero control to the leading subset hierarchy, not by a local boundary instability.  This is compatible with bulk-boundary correspondence, which concerns the leading subset in the TDL, while experimentally accessible samples can display a tunable edge-localized eigenvalue.  The subset windings $\nu_{S_i}$, the finite-$L$ balance of $C_0$ and $C_1$, and weak-coupling flux spectral flow track this distinction.  At critical contact, the ephemeral BIC has a markedly different flux response from the surrounding continuum states.

The observable consequence is scale-dependent spectral separation and boundary localization, whose mechanism's anatomy can be probed via Green's functions.  For $L<L_c$, the OBC spectrum at finite $L$ contains an isolated, strongly localized mode near $E_\ast$.  Across $L_c$, this mode no longer appears as an isolated eigenenergy, although the subleading proj-GF zero remains subset-resolved.  The eigenvalues and left/right eigenvectors needed for the edge density or bi-IPR in Fig.~\ref{fig:main_biIPR} can in principle be retrieved from measured responses using Green's-function methods~\cite{ZhongKimChenLuDingJing2025GreenFunction}.  While the mechanism is completely platform-independent, suitable physical setups are most easily constructed in photonic, topolectrical, acoustic, active-mechanical, and related non-Hermitian platforms~\cite{Longhi2019Probing,song2020two,zhu2020photonic,Xiao2020QuantumDynamics,Helbig2020Topolectrical,Hofmann2020ReciprocalSkin,liu2021non,zou2021observation,zhu2023higher,ZhangYangGe2021Acoustic,Ghatak2020ActiveMechanical,Brandenbourger2019Robotic,liang2022dynamic,shen2023observation}.  

\noindent\textcolor{darkgreen}{\textit{Acknowledgements --}} S.-Y. Y. thanks Albrecht Böttcher for helpful discussions.

\bibliography{references}

\end{document}